# Spatiotemporally Controlled Room Temperature Exciton Transport under Dynamic Pressure

*Kanak Datta[1], Zhengyang Lyu[2], Zidong Li[1], Takashi Taniguchi[3], Kenji Watanabe[4], Parag B. Deotare[1*].*

[1] Department of Electrical and Computer Engineering, College of Engineering, University of Michigan, Ann Arbor, MI -48109, USA.

[2] Applied Physics Program, College of Literature, Science, and Arts, University of Michigan, Ann Arbor, MI - 48109, USA

[3] International Center for Materials Nanoarchitectonics, National Institute for Materials Science, 1-1 Namiki, Ibaraki 305-0044, Japan.

[4] Research Center for Functional Materials, National Institute for Materials Science, 1-1 Namiki, Tsukuba 305-0044, Japan.

ABSTRACT

Two-dimensional transition metal dichalcogenides (TMDs) provide an attractive platform for studying strain dependent exciton transport at room temperature due to large exciton binding energy and strong bandgap sensitivity to mechanical stimuli. Here, we use Rayleigh type surface acoustic wave (SAW) to demonstrate controlled and directional exciton transport under weak coupling regime at room temperature. We screen the in-plane piezoelectric field using photogenerated carriers to study transport under type-I bandgap modulation and measure a maximum exciton drift velocity of 600 m/s. Furthermore, we demonstrate precise steering of exciton flux by controlling the relative phase between the input RF excitation and exciton photogenerations. The results provide important insight into the weak coupling regime between dynamic strain wave and room temperature excitons in a 2D semiconductor system and pave way to exciting applications of excitonic devices in data communication and processing, sensing and energy conversion.

INTRODUCTION:

Moore's law[1] has been empirically describing the scaling of electronics for over half a century. With questionable sustenance for the trend to continue for the next score, various alternative pathways are being explored for next generation data processing and communication technologies. Bound electron-hole pair known as an exciton, provide a viable solution due to its charge neutral nature (free of parasitic capacitance), small dimension, and seamless transition with a photon (ideal for long distance communication links). While the use of excitons for information processing and communication has been proposed for over a decade,[2–4] controlled spatial manipulation of exciton fluxes at room temperature has been challenging, primarily limited by the material system. With the recent emergence of 2D semiconductors such as transition metal dichalcogenides (TMDs) that support excitons with high diffusivity and binding energy (>100 meV)[5,6], feasibility of room temperature excitonic devices is no longer questionable[7]. Unlike charged particles that drift under externally applied electric field, directed transport of charge neutral exciton flux is achieved by spatial tuning of exciton potential by external stimuli such as mechanical strain [8–11] or electric field [7,12–18]. Travelling surface waves (SAWs)[19] can utilize both the effects dynamically to achieve long-range transport. Most of the reported SAW-assisted transport is based on transporting the individual charges following exciton dissociation by the piezoelectric field. Transport under dynamic strain from SAWs has also been achieved in III-V quantum well system.[20–22] However it is limited to cryogenic temperatures and utilizes indirect excitons that are created using an external electrical field. With large sensitivity of the bandgap to external strain[23–25], TMDs are well suited to achieve room temperature, directional transport of direct excitons, solely under dynamic strain.

In this work, we study spatiotemporal control of exciton flux in a monolayer tungsten diselenide ($WSe_2$) system at room temperature. High frequency (~745 MHz resonance frequency) Rayleigh type SAWs are generated in a piezoelectric $128^0$ Y-cut lithium niobate substrate ($LiNbO_3$) using interdigitated electrodes (IDTs) (please refer supporting information section 1 for acoustic response of the device). Mechanically exfoliated monolayer $WSe_2$ encapsulated in hexagonal boron nitride (h-BN) was transferred on the SAW delay line using a dry transfer technique. Using phase-synchronized spatiotemporal measurements[26] and utilizing photogenerated free carriers to screen the in-plane electric field, we report directed exciton transport under type-I bandgap modulation. Based on the experiments, we extract drift velocity of 600 m/s for a strain of ~0. 086% and, estimate the neutral exciton mobility to be about 900 $cm^2/eV/s$. In addition, we demonstrate precise manipulation over the exciton transport by controlling the phase delay between RF excitation and photoexcitation. The work provides important insight into the weak coupling regime between dynamic strain wave and room temperature excitons in a 2D semiconductor system, leading to a potential pathway for long range exciton transport at room temperature.

## RESULTS AND DISCUSSION:

Figure 1(a) shows a schematic of the device geometry used in this work. Figure 1(b) and (c) shows a false color brightfield optical image of the sample and the photoluminescence (PL) map of the h-BN encapsulated monolayer $WSe_2$. The h-BN encapsulation is critical to improve the transport properties of the excitons by suppressing non-radiative recombination processes[27], surface roughness[28], energetic disorder[29] and scattering from impurities and surface states[30,31]. More importantly, the underlying bulk h-BN moderates the dielectric environment surrounding the monolayer thereby increasing the exciton binding energy when compared to monolayer directly placed on $LiNbO_3$ substrate. This reduces exciton dissociation under SAW piezoelectric field (type-II modulation) due to increased binding energy.[32–34] The dissociation can be further reduced by screening the piezoelectric field, thereby enabling the study of excitonic interactions with type-I modulation (bandgap change due to strain). In this work, we achieve it by utilizing the optically generated free carriers to screen the in-plane electric field of the traveling SAW wave. Figure 1(d) plots the PL quenching ($r_Q = \frac{I_{PL,RF}}{I_{PL,-60dBm}}$) due to dissociation as a function of RF input power for various optical excitations. Here, $I_{PL,RF} = \int I_{RF}(\lambda)\, d\lambda$ and $I_{PL,-60dBm} = \int I_{-60dBm}(\lambda)\, d\lambda$ refer to integrated PL intensity measured at a given RF input power and at -60 dBm (1 nW, the minimum RF input power used in this work), respectively. The net quenching at a given RF power, reduces with increase in optical excitation density due to optically generated free carrier screening [34,35] (Please refer to supporting information section 2 for estimation of optical excitation density and section 3 for PL spectra). Thus, the excitation optical fluence provides a knob to control the dissociation and thereby investigate effects of type-I modulation on exciton transport.

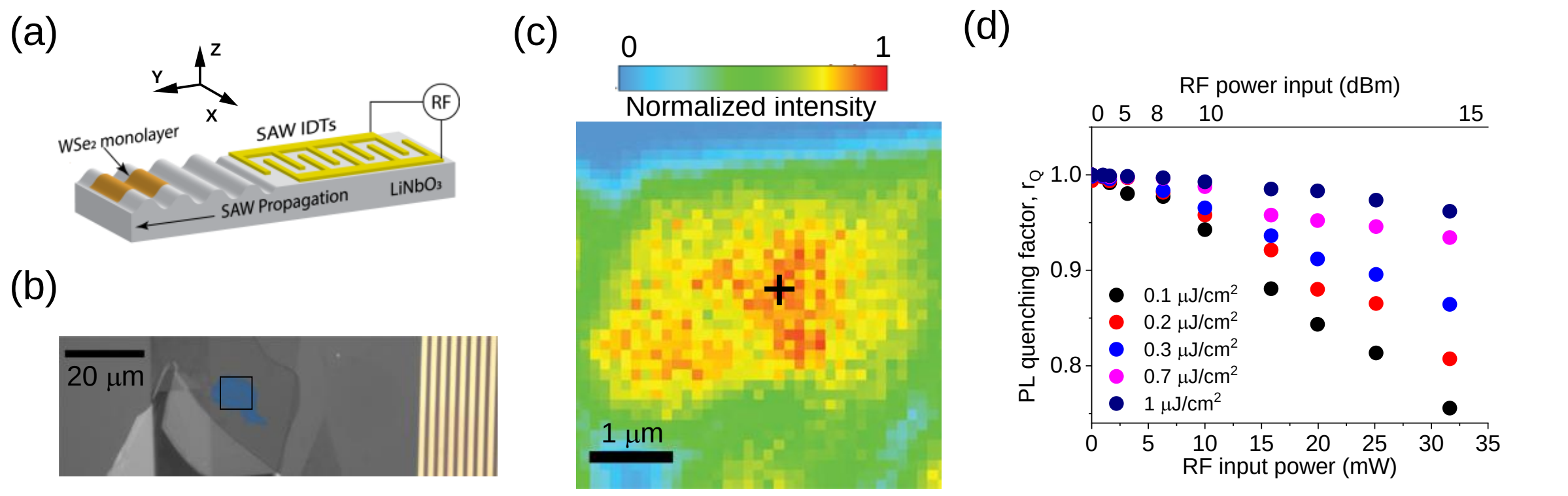

Figure 1: Modulation of h-BN encapsulated monolayer $WSe_2$ photoluminescence (PL) by traveling piezoelectric field of the surface acoustic wave. (a) A schematic representation of the transferred monolayer $WSe_2$ on the SAW delay line. (b) Brightfield optical image of a h-BN encapsulated monolayer $WSe_2$ transferred on lithium niobate substrate. (c) Integrated PL intensity map of monolayer area represented by the square in (b). (d) PL intensity modulation under increased RF power at various optical excitation densities. The monolayer PL emission decreases with increase in RF power. The net PL quenching, however, reduces with increasing optical power density due to screening from optically generated free carriers.

**Acoustic wave mediated exciton transport in monolayer $WSe_2$:**

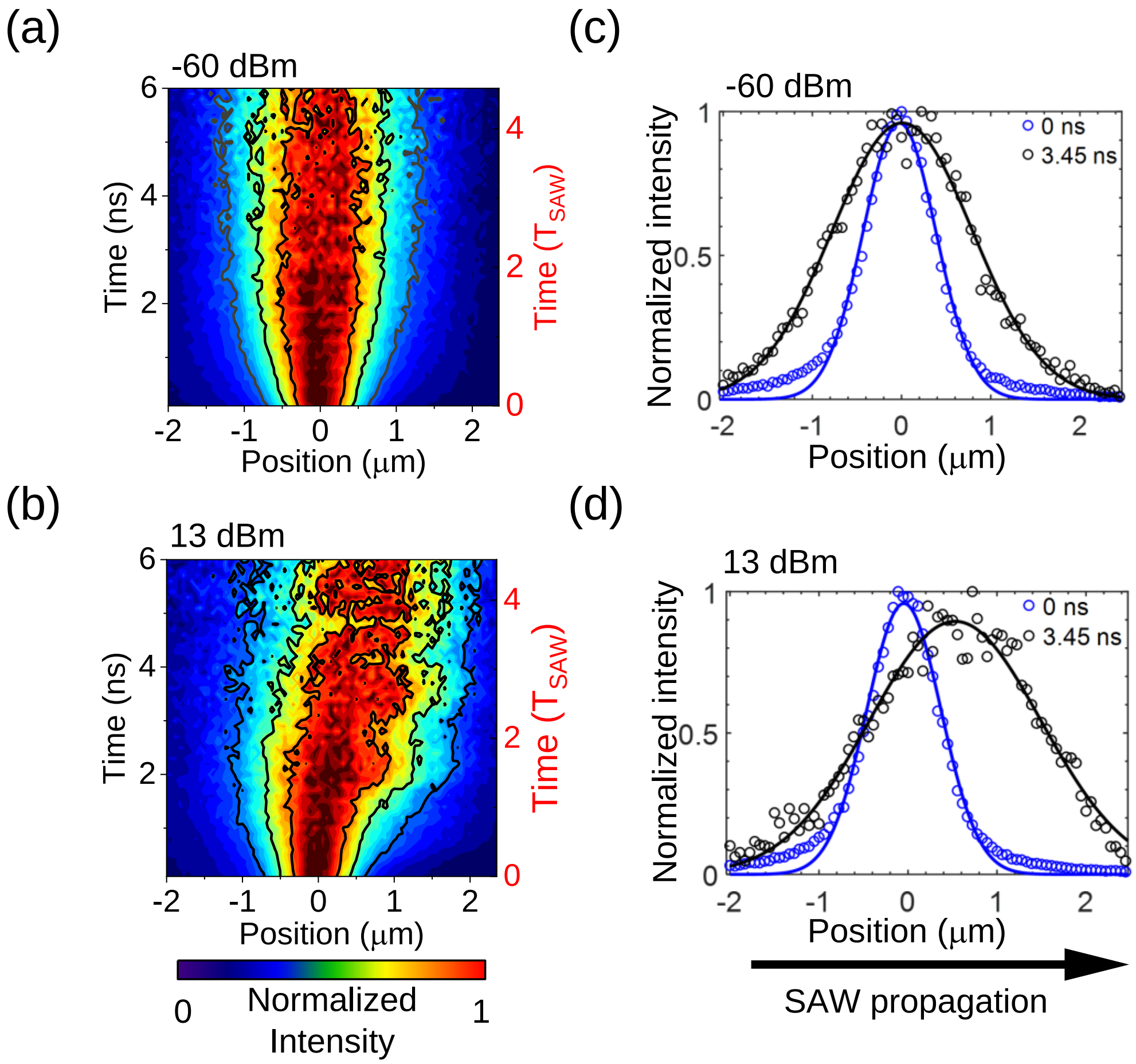

Figure 2: Spatiotemporal map of exciton density (normalized for each time instance along space) for (a) -60 dBm and (b) 13 dBm RF input power at optical fluence of 1.2 μJ/cm$^2$. A spatial shift in exciton density can be observed under RF excitation. Normalized exciton density profiles at different time instances as a function of space along with the Gaussian fit (c) under -60 dBm acoustic power, and (d) 13 dBm RF input power. At minimum RF input power (-60 dBm), the exciton density broadens symmetrically in space due to exciton diffusion in the monolayer. At high RF input, an asymmetric spatial shift of the exciton density in the direction of the acoustic wave propagation is observed with the observed shift increasing with time.

Figure 2 shows the results of excitonic energy transport measured using a scanning SPAD technique described in our previous work[8,26]. Figure 2(a) and (b) show the normalized spatiotemporal exciton density distributions at -60 dBm (~1 nW) and 13 dBm (~20 mW) RF input power respectively at an optical fluence of 1.2 μJ/cm$^2$ (PL quenching is ~2 % under these conditions and hence type-I modulation dominates). The exciton density distributions have been normalized for each time instance. Figure 2(c) and (d) show the normalized exciton density distribution along with the Gaussian fit at two-time instances extracted from Figure 2 (a) and (b) respectively. The symmetric exciton density distribution at -60dBm resembles typical anomalous exciton diffusion in monolayer $WSe_2$ (refer supporting information section 4 for related information on extraction of diffusivity). In this case, the peak position of the distribution does not show any spatial drift with time, indicating the absence of local strain gradient in the monolayer. When the RF excitation is turned ON, the exciton density distribution shifts along the propagation direction (black arrow) of the SAW wave as shown in Figure 2 (d).

Figure 3(a) shows the spatial evolution of the gaussian exciton density peak as a function of time. With increasing RF power, we see a gradual increase in the spatial shift. In addition, we observe periodic oscillations corresponding to the SAW period. This suggests weak coupling regime between the excitons and the traveling strain wave (discussed in the following section), where the exciton drift velocity is insufficient to keep up with the travelling wave and hence, results in a net spatial shift over time of the exciton density. We model the evolution of the gaussian peak under dynamic strain over time using a linear relationship (solid line in Figure 3(a)):

$$h(t) = v_{avg}t + h_{offset} \tag{1}$$

where, $h(t)$ refers to the net displacement of the generated exciton density over many SAW periods, $v_{avg}$ refers to the average drift velocity of the exciton density under applied dynamic strain, and $h_{offset}$ is a constant to accommodate the fitting error. Figure 3(b) plots the $v_{avg}$ for various volumetric strain estimated at different RF input power. A linear trend with volumetric strain indicates a proportional increase in exciton coupling efficiency with the dynamic strain field. Further improvement in $v_{avg}$ can be achieved by increasing the strain gradient or increasing exciton diffusivity by suppressing scattering from impurity states and surface roughness[36].

We determine the extent of exciton coupling to the dynamically varying strain field of the SAW by estimating the instantaneous velocity from the measured instantaneous displacement of the exciton density. The time derivative of the measured displacement at 13 dBm RF input power gives the instantaneous velocity of the exciton flux as shown in Figure 3(c) The instantaneous drift velocity reaches a maximum value of 600 m/s, which is smaller than the SAW velocity ($v_{saw}$) in $LiNbO_3$ (3979 m/s[37]). Since the maximum drift velocity of the exciton flux is nearly six times smaller than the acoustic wave velocity ($v_{max,13\ dBm} < v_{saw}$), the exciton flux cannot keep pace with the traveling strain wave (weak coupling regime). Hence, we observe oscillations in the exciton flux and a net spatial shift over time in the direction of the traveling strain field as seen in Figure 2(b). This observation closely resembles carrier drift under dynamic electric field in the weak coupling regime presented in García-Cristóbal et al.[38].

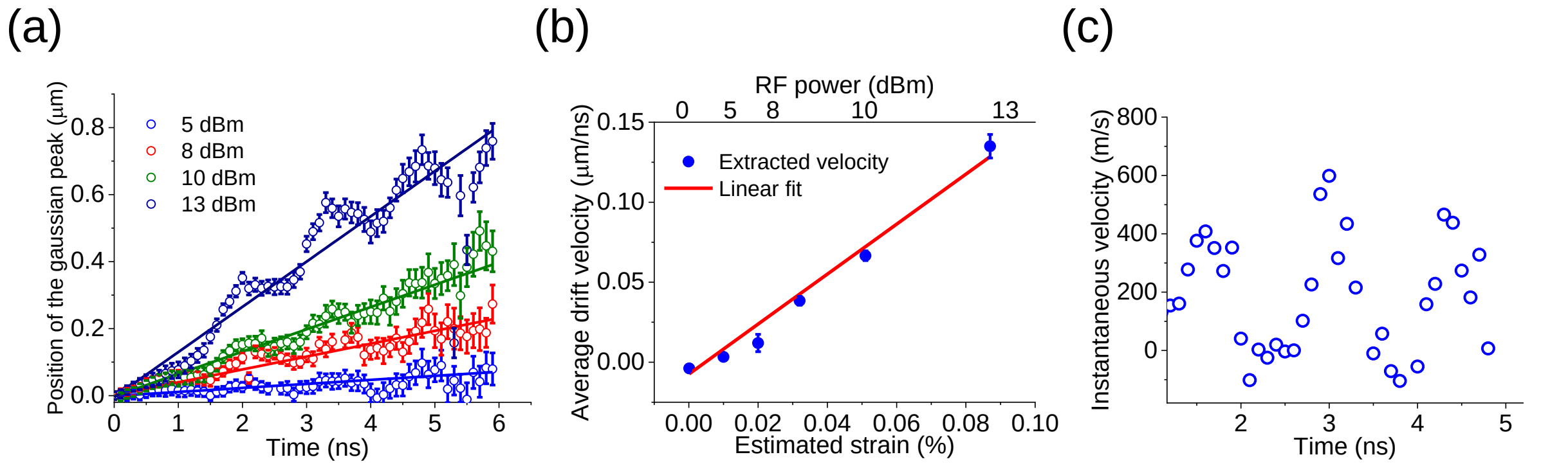

Figure 3: Exciton transport in monolayer $WSe_2$ under incremental RF power at 1.2 μJ/cm$^2$ optical fluence. (a) Spatiotemporal evolution of the exciton density peak extracted using gaussian fits (circles) and the approximate linear fit (solid lines) at various acoustic power. The error bars represent the uncertainties in the gaussian fit. (b) Effective shift in the position of the gaussian peak extracted from the model introduced in the main text that shows an increase in transport length as RF input power is increased. The maximum shift of the Gaussian peak increases with applied RF input power, clear signature of long-range energy transport under dynamic strain in the monolayer. (c) The time derivative of the 13 dBm data in (a). The instantaneous velocity of the exciton flux reaches 600 m/s but is still lower than the acoustic wave velocity (3979 m/s).

Under a dynamically varying strain field, the maximum exciton drift velocity can be written as:

$$v_{max} = \mu \left|\frac{\partial E_g}{\partial \varepsilon}\right| \frac{\partial \varepsilon}{\partial x}|_{max}$$

$$= \mu \left|\frac{\partial E_g}{\partial \varepsilon}\right| k\varepsilon_0 \qquad (2)$$

where, $\mu$, $\frac{\partial E_g}{\partial \varepsilon}$, $k = \frac{2\pi}{\lambda_{SAW}}$, and $\varepsilon_0$ refer to the exciton mobility, strain sensitivity of monolayer bandgap,[39] acoustic wave momentum and maximum dynamic strain in the monolayer, respectively. $\lambda_{SAW}$ refers to the wavelength of the acoustic wave ($\lambda_{SAW} = 4.7\ \mu m$). Based on the measured PL quenching ( Figure 1(d)), we calculate the maximum dynamic strain amplitude ($\varepsilon_0$) to be ~0.086 % at 13 dBm RF power using the converse piezoelectric matrix[40] of $128^0$ $LiNbO_3$ and the estimated piezoelectric field in the substrate[41] (refer supporting information section 10 for details). Using the estimated value of maximum drift velocity at 13 dBm RF input power in equation (2), we extract the exciton mobility in the monolayer to be 900 $cm^2/eV/s$. The extracted exciton mobility is nearly two orders lower (exceeding $10^4$ $cm^2/V/s$ [42–45]) compared to indirect excitons in III-V quantum well structures at cryogenic temperatures, and thus results in weak coupling.

**Acoustic steering of photogenerated exciton density:**

The dynamic acoustic strain field generated by a SAW wave can be represented by:

$$\varepsilon(x,t) = \varepsilon_0 \cos(\omega t - kx + \varphi) \qquad (3)$$

where, $\varepsilon_0$ refers to the amplitude of the strain field, $\omega$ refers to the angular frequency ($\omega = 2\pi f_{SAW}$; $f_{SAW}$refers to the acoustic resonance frequency), and $\varphi$ refers to the instantaneous phase of the acoustic wave. Therefore, for a given optical excitation position from the IDT, the generated exciton density interacts with a certain phase of the traveling strain field. Using equation (3), the dynamic strain induced drift velocity of the exciton flux can be written as[8]:

$$v_{ex}(x,t) = \mu_{\varepsilon}\frac{\partial \varepsilon(x,t)}{\partial x} = \mu_{\varepsilon} k\, sin(\,\omega t - kx + \varphi) \qquad (4)$$

where $\mu_{\varepsilon} = \mu \frac{\partial E_g}{\partial \varepsilon}$ refers to the strain mobility of the excitons[8,46]. At photoexcitation (t=0), the instantaneous velocity of the exciton flux is a function of the acoustic phase, $v_{ex}(x,0) = \mu_{\varepsilon} k\, sin(-kx + \varphi)$. Therefore, precise control over the direction of photogenerated exciton flux can be achieved by controlling the relative phases as shown in Figure 4(a) – (d). We achieve such control by introducing a delay, τ ($\varphi = \frac{2\pi}{T}\tau; T = \frac{1}{f_{SAW}}$) in the laser trigger signal (refer supporting material section 5 for schematics of the characterization setup). Figure 4(e) plots the spatiotemporal evolution of the center of the exciton density distribution as a function of the time delay (τ) with increments of T/4 (refer section 6 of supporting information for the corresponding spatiotemporal exciton density maps), which were further numerically modeled using the following modified drift-diffusion equation[8,46]:

$$\frac{\partial n_{ex}(x,t)}{\partial t} = D_{ex}\frac{\partial^2 n_{ex}(x,t)}{\partial x^2} + \frac{\partial (v_{ex} n_{ex}(x,t))}{\partial x} + G - \frac{n_{ex}(x,t)}{\tau_{ex}} - \frac{n_{ex}(x,t)}{\tau_{ion}} \qquad (5)$$

Here, $n_{ex}$, $D_{ex}$, $v_{ex}$, $\tau_{ex}$ and $\tau_{ion}$ refer to the population, diffusion co-efficient, velocity, recombination time and ionization time of the neutral exciton[38] respectively. $G$ refers to the photogenerated exciton density under optical excitation. The solid lines in Figure 4(e) are fits from the numerical simulation that successfully captures the observed dynamics (refer supporting information section 7 for details on the formulation). Consistently, we observed similar progressive evolution with instantaneous phase in the TRPL (refer supporting information section 9 for experimental data).

Exciton transport under type-I modulation takes place by spatial trapping at the minimum energy locations[19,21](please refer Figure 4(a)-(d) for schematic representation for the energy band modulation under dynamic strain field). We verified that the observed spatiotemporal modulation in the exciton density is due to type-I modulation by performing three experiments. First, we conduct transport measurements at an order lower optical fluence. Since the effective mobility of the photogenerated excitons increase with optical fluence[47], the excitons experience efficient coupling with the traveling strain field at high optical excitation densities. For the same set of RF powers (constant strain), we observed reduced modulation in the exciton density and a decrease in average exciton spatial drift for lower optical fluence (refer to supporting information section 8 for experiments and detailed discussion). Second, we measured the time dependent photoluminescence (TRPL) as a function of optical fluence. At low fluence, the TRPL drops faster with increase in RF power due to increased ionization (additional decay) resulting from type-II modulation. At high fluence, the TRPL recovers with increased RF power, confirming screening of the in-plane piezoelectric field by the free carriers (refer Figure 1(d)) (refer to supporting information section 9 for TRPL experiments). However, TRPL oscillates at the SAW frequency. The strain induced oscillations result from the dynamic modulation of the energy separation between K and Q valleys under the traveling strain wave[9,24] (refer to supporting information section 9 for the model describing the strain induced oscillations). Finally, we verify type–I band edge modulation by extracting the neutral exciton linewidth broadening, which remains constant for varying optical excitation fluence under RF excitation (refer to supporting information section 10 for PL broadening under type-I and type-II bandgap modulations).

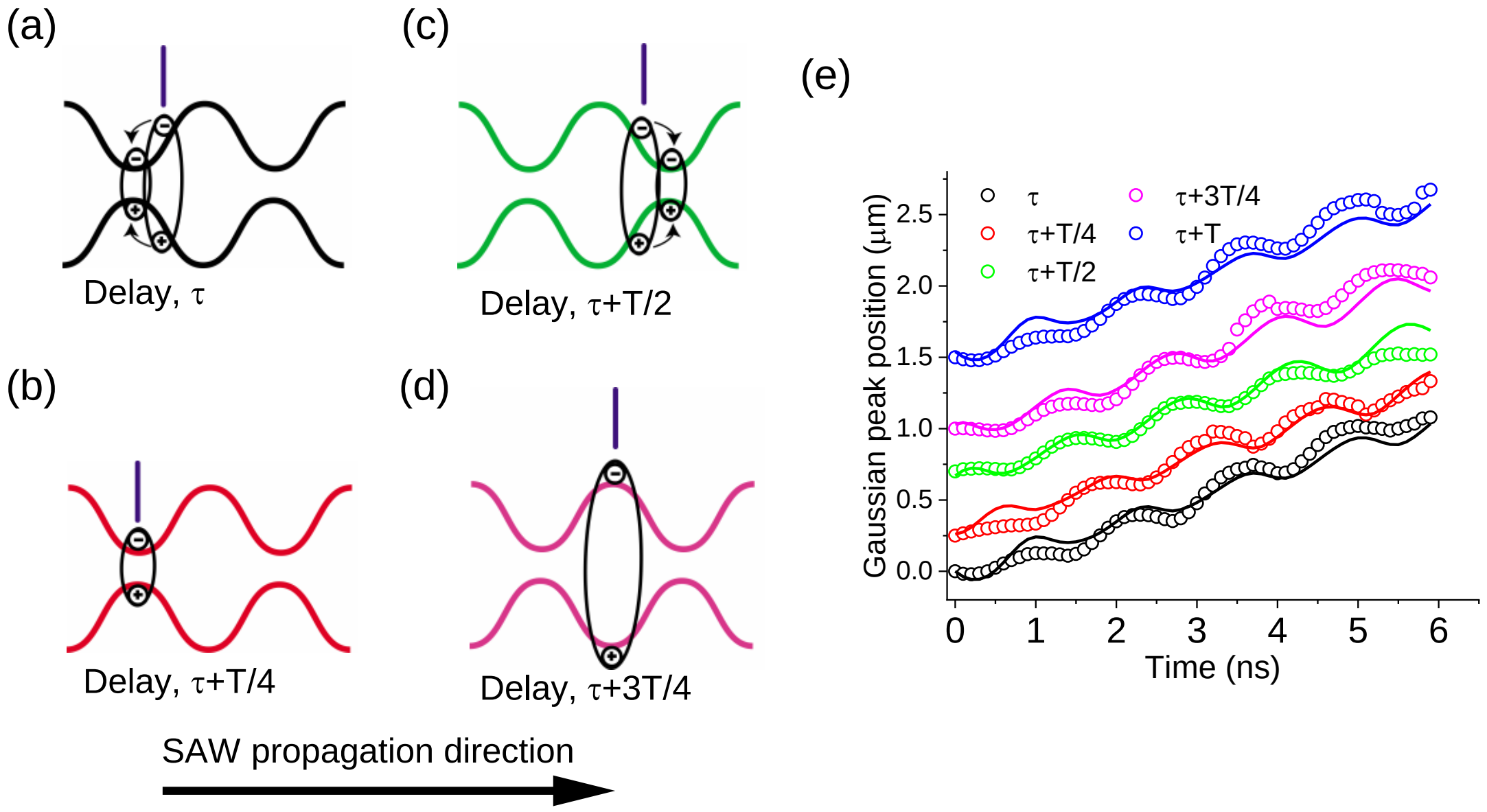

Figure 4: Spatiotemporally controlled energy transport under phase modulated optical excitation for 15 dBm RF excitation. Schematic presentation of photogenerated excitons at different SAW phases: (a) τ; (b) τ + T/4; (c) τ + T/2; (d) τ + 3T/4. The vertical blue line represents the position of excitation pulse on SAW wave. For τ, the excitons are generated at a position such that the energetically stable position lies in the opposite direction of SAW propagation. Therefore, immediately after photogeneration, the exciton density distribution shifts in the opposite direction of SAW propagation. However, as the excitons couple to the traveling strain field, a net drift in the direction of SAW was observed. For τ+T/4, excitons are generated at the positions of lowest bandgap (position of maximum tensile strain corresponding to minimum exciton potential but lowest energy gradient). Hence, the exciton density distribution does not undergo immediate spatial shift after photogeneration. At τ+T/2, following photoexcitation, the energy gradient drives the exciton flux forward towards the stable position of maximum tensile strain. Therefore, the generated density distribution moves immediately in the direction of SAW propagation. For τ+3T/4, the excitons are generated at the position of maximum compressive strain *i.e.*, maximum bandgap and hence results in no spatial shift immediately after photogeneration. (e)Phase synchronized evolution of the gaussian peak position(circles) extracted by fitting the normalized exciton density measured using TCSPC technique. The gaussian peak position oscillates at the period of the SAW wave. A progressive delay in the Gaussian peak is also observed as the time delay is increased at an increment of T/4. Alongside oscillations, a net drift in the Gaussian peak is observed in the direction of acoustic wave propagation. The solid line in the figure shows numerical fit using the model discussed in supporting information section 7.

In conclusion, we demonstrated room temperature directed exciton energy transport in monolayer $WSe_2$ under a traveling strain field. The dynamic strain gradient obtained under the experimental condition was sufficient to study the weak coupling regime between excitons and the strain wave. Our results show that the weak coupling leads to oscillations in the transported exciton density since the exciton drift velocity is lower than the velocity of the travelling wave. Based on the measurement, we estimate the neutral exciton mobility in the monolayer to be 900 $cm^2/eV/s$ which is in good agreement with the values reported in literature for the same material. The results show that coupling between excitons and dynamic strain wave in TMDs strongly depends on the transport properties and therefore, is expected to be influenced by various factors like intrinsic and extrinsic defect states, lattice disorders, substrate roughness, scattering from charged and neutral impurities. Hence, improved material system with higher strain gradients would provide a pathway to reach the strong coupling regime. In addition, the use of TMD heterostructures that offer long lifetime could aid in improving the overall transport length. Finally, we also demonstrated acoustic steering over the photogenerated excitons by precisely tuning the exciton photogeneration with respect to the phase of the acoustic wave and provide a modified drift-diffusion model to the describe the observations. The reported results pave the way for exciting future applications that include efficient energy conversion, sensing, detection and room temperature on-chip excitonic information processing and communication.

Methods:

SAW Device Fabrication:

The SAW devices were fabricated using standard photolithography-based metal lift-off process on a $128^0$ Y-cut lithium niobate substrate ($LiNbO_3$). A bi-layer photoresist stack was used (LOR 3A+ S1813) to achieve high resolution IDT features. The IDTs were patterned using a projection lithography tool (GCA Autostep AS200). Following development of the exposed features, 10 nm Cr and 100 nm Au was evaporated using electron beam evaporation technique. The lift-off was performed by immersing the samples in Remover PG for 12 hours. A second step lithography was performed to pattern the contact pads. The second step lithography was carried out using SPR 220(3.0) as the photoresist layer. The same projection lithography tool was used to pattern the contact pads. Following development of the exposed contact pad features, 10 nm Cr and 480 nm Au was evaporated using electron beam evaporation technique. The lift-off was performed following the same process as step 1. The samples were bonded to a custom-made PCB board using wire bonding.

Detailed parameters for the overall process of device fabrication can be found in our previous work. Detailed parameters for the overall process of device fabrication can be found in our previous work[34].

Exfoliation and Transfer of Monolayer $WSe_2$:

The h-BN and $WSe_2$ flakes were mechanically exfoliated from bulk crystals (bulk h-BN crystal was received from Takashi Taniguchi and Kenji Watanabe. Bulk $WSe_2$ was purchased from HQ Graphene) using scotch tapes. The top/bottom h-BN and the monolayer $WSe_2$ were then exfoliated from the scotch tapes onto a thin polydimethylsiloxane (PDMS) stamp (GEL-FILM WF-40/1.5-X4) and identified under an optical microscope. Heterostructures were stacked layer by layer using a polymer-assisted dry transfer technique[48] under a home-built setup with micromanipulators and a rotational stage. No heat was applied during the transfer process.

Acoustic Response Measurement from the SAW Devices:

The acoustic characterization of the SAW devices was carried out using a network analyzer (Agilent 4396B) connected to a S-parameter test unit (Agilent 85046A). The results of acoustic response measurement can be found in the supporting information section 1.

Exciton transport Measurement Setup:

The optical characterization was carried out using a phase synchronized time correlated single photon counting (TCSPC) setup. The sample was non-resonantly excited (Excitation wavelength 405 nm) using a pulsed diode(laser diode source - PicoQuant LDH P-C-405 and laser diode driver - PDL 800-D) laser producing pulses ~30 ps in duration. The photoluminescence signal was collected using a highly sensitive avalanche photodiode (MPD PDM series[49]) through a 60X dry objective (NIKON CFI PLAN APO λ 60X / 0.95). The output of the APD was analyzed with a timing module (PicoQuant HydraHarp 400) of ~1 ps time resolution. The TCSPC setup was phase synchronized with a vector network analyzer (Agilent/HP 4396B - used as the RF signal source) using a variable pulsed delay generator(Stanford Research Systems DG645). To achieve complete phase synchronization, the output repetition rate from the delay generator was fixed at an integer multiple of the time-period of the RF signal. The output pulse stream (pulse width ~ 20 ns) from the delay generator was used to trigger the picosecond laser diode module. A complete schematic representation of the phase synchronized setup can be found in the supporting information section 5.

CW PL measurement setup:

The CW PL was measured using a 450 nm CW laser using a diffraction limited spot. The PL was collected using a 40X 0.95 NA objective (Nikon Plan Apo 40X 0.95) and analyzed using a high-resolution spectrometer (Princeton Instruments IsoPlane SCT 320) coupled to a highly sensitive CCD camera (Princeton Instruments, Model Pixis 400).

**Corresponding Author**

* Parag B. Deotare − Department of Electrical and Computer Engineering, University of Michigan, Ann Arbor, Michigan 48109, United States

**Author Contributions**

PBD conceived the idea and supervised the project. KD fabricated and characterized the devices. ZY transferred the encapsulated monolayers on SAW devices. PBD, KD, and ZL analyzed the data. Growth of the h-BN was done by TT and KW. All authors contributed in writing the manuscript.

**Funding Sources**

PBD acknowledges partial support of this work by the Air Force Office of Scientific Research (AFOSR) award no. FA9550-17-1-0208 and by the Army Research Office under Grant Number W911NF-21-1-0207. K.W. and T.T. acknowledge support from the Elemental Strategy Initiative conducted by the MEXT, Japan (Grant Number JPMXP0112101001) and JSPS KAKENHI (Grant Numbers 19H05790 and JP20H00354).

ACKNOWLEDGMENT

The authors acknowledge the help and support from the Lurie Nanofabrication Facility at the University of Michigan, Ann Arbor where the device fabrication was carried out.

REFERENCES

1. Waldrop, M. M. The chips are down for Moore's law. *Nature* **530**, 144–147 (2016).

2. Grosso, G. *et al.* Excitonic switches operating at around 100K. *Nat. Photonics* **3**, 577–580 (2009).

3. Butov, L. V. Excitonic devices. *Superlattices Microstruct.* **108**, 2–26 (2017).

4. Baldo, M. & Stojanovi, V. Optical switching: Excitonic interconnects. *Nat. Photonics* **3**, 558–560 (2009).

5. Manzeli, S., Ovchinnikov, D., Pasquier, D., Yazyev, O. V. & Kis, A. 2D transition metal dichalcogenides. *Nat. Rev. Mater.* **2**, 1–15 (2017).

6. Chernikov, A. *et al.* Exciton binding energy and nonhydrogenic Rydberg series in monolayer $WS_2$. *Phys. Rev. Lett.* **113**, 076802 (2014).

7. Fowler-Gerace, L. H., Choksy, D. J. & Butov, L. V. Voltage-controlled long-range propagation of indirect excitons in van der Waals heterostructure. (2021).

8. Cordovilla Leon, D. F., Li, Z., Jang, S. W., Cheng, C. H. & Deotare, P. B. Exciton transport in strained monolayer $WSe_2$. *Appl. Phys. Lett.* **113**, 252101 (2018).

9. Moon, H. *et al.* Dynamic exciton funneling by local strain control in a monolayer semiconductor. *Nano Lett.* **20**, 6791–6797 (2020).

10. Palacios-Berraquero, C. *et al.* Large-scale quantum-emitter arrays in atomically thin semiconductors. *Nat. Commun.* (2017). doi:10.1038/ncomms15093

11. Branny, A., Kumar, S., Proux, R. & Gerardot, B. D. Deterministic strain-induced arrays of

quantum emitters in a two-dimensional semiconductor. *Nat. Commun.* **8**, 1–7 (2017).

12. Unuchek, D. *et al.* Room-temperature electrical control of exciton flux in a van der Waals heterostructure. *Nature* **560**, 340–344 (2018).

13. High, A. A., Novitskaya, E. E., Butov, L. V., Hanson, M. & Gossard, A. C. Control of exciton fluxes in an excitonic integrated circuit. *Science (80-. ).* **321**, 229–231 (2008).

14. High, A. A., Hammack, A. T., Butov, L. V., Hanson, M. & Gossard, A. C. Exciton optoelectronic transistor. *Opt. Lett.* **32**, 2466 (2007).

15. Winbow, A. G. *et al.* Electrostatic conveyer for excitons. *Phys. Rev. Lett.* **106**, 1–4 (2011).

16. Liu, Y. *et al.* Electrically controllable router of interlayer excitons. *Sci. Adv.* **6**, eaba1830 (2020).

17. Jiang, Y., Chen, S., Zheng, W., Zheng, B. & Pan, A. Interlayer exciton formation, relaxation, and transport in TMD van der Waals heterostructures. *Light Sci. Appl.* **10**, 2047–7538 (2021).

18. Unuchek, D. *et al.* Valley-polarized exciton currents in a van der Waals heterostructure. *Nat. Nanotechnol.* **14**, 1104–1109 (2019).

19. Rudolph, J., Hey, R. & Santos, P. V. Exciton transport by surface acoustic waves. *Superlattices Microstruct.* **41**, 293–296 (2007).

20. Rudolph, J., Hey, R. & Santos, P. V. Long-range exciton transport by dynamic strain fields in a GaAs quantum well. *Phys. Rev. Lett.* **99**, 1–4 (2007).

21. Violante, A. *et al.* Dynamics of indirect exciton transport by moving acoustic fields. *New J.*

*Phys.* **16**, 033035 (2014).

22. Lazić, S. *et al.* Scalable interconnections for remote indirect exciton systems based on acoustic transport. *Phys. Rev. B - Condens. Matter Mater. Phys.* **89**, 1–8 (2014).

23. Desai, S. B. *et al.* Strain-induced indirect to direct bandgap transition in multilayer $WSe_2$. *Nano Lett.* **14**, 4592–4597 (2014).

24. Aslan, O. B., Deng, M. & Heinz, T. F. Strain tuning of excitons in monolayer $WSe_2$. *Phys. Rev. B* **98**, 115308 (2018).

25. Island, J. O. *et al.* Precise and reversible band gap tuning in single-layer $MoSe_2$ by uniaxial strain. *Nanoscale* **8**, 2589–2593 (2016).

26. Akselrod, G. M. *et al.* Visualization of exciton transport in ordered and disordered molecular solids. *Nat. Commun.* **5**, 1–8 (2014).

27. Hoshi, Y. *et al.* Suppression of exciton-exciton annihilation in tungsten disulfide monolayers encapsulated by hexagonal boron nitrides. *Phys. Rev. B* **95**, 241403 (2017).

28. Dean, C. R. *et al.* Boron nitride substrates for high-quality graphene electronics. *Nat. Nanotechnol.* **5**, 722–726 (2010).

29. Wierzbowski, J. *et al.* Direct exciton emission from atomically thin transition metal dichalcogenide heterostructures near the lifetime limit. *Sci. Rep.* **7**, 1–6 (2017).

30. Morozov, S. V. *et al.* Giant intrinsic carrier mobilities in graphene and its bilayer. *Phys. Rev. Lett.* **100**, 016602 (2008).

31. Hwang, E. H., Adam, S. & Sarma, S. Das. Carrier transport in two-dimensional graphene

layers. *Phys. Rev. Lett.* **98**, 186806 (2007).

32. Lin, Y. *et al.* Dielectric screening of excitons and trions in single-layer $MoS_2$. *Nano Lett.* **14**, 5569–5576 (2014).

33. Pedersen, T. G. Exciton Stark shift and electroabsorption in monolayer transition-metal dichalcogenides. *Phys. Rev. B* **94**, 125424 (2016).

34. Datta, K., Li, Z., Lyu, Z. & Deotare, P. B. Piezoelectric Modulation of Excitonic Properties in Monolayer $WSe_2$ under Strong Dielectric Screening. *ACS Nano* acsnano.1c04269 (2021). doi:10.1021/acsnano.1c04269

35. Santos, P. V., Ramsteiner, M. & Jungnickel, F. Spatially resolved photoluminescence in GaAs surface acoustic wave structures. *Appl. Phys. Lett.* **72**, 2099–2101 (1998).

36. Hotta, T. *et al.* Exciton diffusion in a h BN-encapsulated monolayer $MoSe_2$. *Phys. Rev. B* **102**, 115424 (2020).

37. Morgan, D. *Surface Acoustic Wave Filters*. *Surface Acoustic Wave Filters* (Elsevier Ltd, 2007). doi:10.1016/B978-0-12-372537-0.X5000-6

38. García-Cristóbal, A., Cantarero, A., Alsina, F. & Santos, P. V. Spatiotemporal carrier dynamics in quantum wells under surface acoustic waves. *Phys. Rev. B - Condens. Matter Mater. Phys.* **69**, 1–13 (2004).

39. Aas, S. & Bulutay, C. Strain dependence of photoluminescence and circular dichroism in transition metal dichalcogenides: a *k.p* analysis. *Opt. Express* **26**, 28672 (2018).

40. Shur, V. Y. Lithium niobate and lithium tantalate-based piezoelectric materials. in

*Advanced Piezoelectric Materials: Science and Technology* 204–238 (Elsevier Inc., 2010). doi:10.1533/9781845699758.1.204

41. Lazić, S. *et al.* Dynamically tuned non-classical light emission from atomic defects in hexagonal boron nitride. *Commun. Phys.* **2**, 1–8 (2019).

42. Hillmer, H. *et al.* Optical investigations on the mobility of two-dimensional excitons in GaAs/$Ga_{1-x}$ $Al_xAs$ quantum wells. *Phys. Rev. B* **39**, 10901 (1989).

43. Dorow, C. J. *et al.* High-mobility indirect excitons in wide single quantum well. *Appl. Phys. Lett.* **113**, 212102 (2018).

44. Bacher, G. *et al.* Correlation between the exciton mobility and the excitonic linewidth in shallow $In_xGa_{1-x}As$/GaAs quantum wells. *Appl. Phys. Lett.* **61**, 702–704 (1992).

45. Gärtner, A., Holleitner, A. W., Kotthaus, J. P. & Schuh, D. Drift mobility of long-living excitons in coupled GaAs quantum wells. *Appl. Phys. Lett.* **89**, 52108 (2006).

46. Fu, X. *et al.* Exciton drift in semiconductors under uniform strain gradients: Application to bent ZnO microwires. *ACS Nano* **8**, 3412–3420 (2014).

47. Perea-Causín, R. *et al.* Exciton Propagation and Halo Formation in Two-Dimensional Materials. *Nano Lett.* **19**, 7317–7323 (2019).

48. Castellanos-Gomez, A. *et al.* Deterministic transfer of two-dimensional materials by all-dry viscoelastic stamping. *2D Mater.* **1**, 011002 (2014).

49. Micro Photon Devices - PDM. Available at: http://www.micro-photon-devices.com/Products/SPAD-by-Wavelength/400nm-900nm/PDM-PDF.